\documentclass[]{aa}

\usepackage{graphicx}

\usepackage{txfonts}

\usepackage{color}

\usepackage{natbib}

\usepackage{color}

\newcommand*{\dd}{{\textrm d}}                       
\newcommand*{\g}{{\mbox{\textsl g}}}                 
    
\newcommand*{\pd}{\partial}                          
\newcommand*{\pD}[2]{\frac{\pd{#1}}{\pd{#2}}}        
\newcommand{\non}{\nonumber}

\newcommand*{\T}{\textstyle}

\def\sss{\scriptscriptstyle}
\def\U{{\sss \!U}}
\def\L{{\sss \!L}}
\def\K{{\sss \!K}}
\def\LT{{\sss \!L\,\!T}}

\def\T{{\sss \!T}}

\def\P{{\sss \!P}}
\def\nur{\nu_\mathrm{r}}
\def\nuv{\nu_\theta}
\def\nuL{\nu_\L}
\def\nuU{\nu_\U}
\def\nuK{\nu_\K}
\def\nuLT{\nu_\LT}
\def\nul{\nuL}
\def\nuu{\nuU}
\def\nuBTS{\nu_\T}
\def\nuP{\nu_\P}
\def\dcp{total precession frequency}
\begin{document}

\title{On a multi-resonant origin of high frequency quasiperiodic oscillations in the neutron-star X-ray binary \mbox{4U~1636-53}}

\author
{Zden\v{e}k Stuchl\'{\i}k
\and
Gabriel T\"or\"ok
\and
Pavel Bakala
}

\offprints{G.~T\"or\"ok, ~ terek@volny.cz}

\institute{Institute of Physics, Faculty of Philosophy and Science, Silesian
  University in Opava,\\ Bezru\v{c}ovo n\'{a}m. 13,CZ-74601 Opava, Czech Republic}
 
\date{Received / Accepted}

\abstract
{
The kHz quasiperiodic oscillations (QPOs) observed in low-mass X-ray neutron star binaries are most likely connected to the orbital motion in the accretion disc. The ratio between frequencies of the upper and lower observed QPOs mode cluster close to ratios of small natural numbers, most often close to the $3/2$ value, but the other rational ratios occur in some sources as well. The class of QPOs models considers a resonance between Keplerian and epicyclic frequencies of the geodesic motion. It was suggested that a multipeaked ratio distribution may follow from different resonances. The atoll source \mbox{4U~1636-53} shows datapoints clustering around two distinct values ($3/2$ and $5/4$) of the frequency ratio. The same frequency ratios correspond to the change in the sign of the twin peak QPOs amplitude difference, suggesting existence of a resonant energy overflow.}
{We explore the idea that the two clusters of datapoints in 4U 1636-53 result from two different instances of the same orbital resonance corresponding to the two resonant points.} 
{Assuming the neutron star external spacetime to be described by the Hartle-Thorne metric, we search for a frequency relation, matching the two observed datapoints clusters, which may correspond to a resonance. We consider orbital and associated epicyclic frequencies with accuracy up to the second order terms in the neutron star angular momentum $j$ and first order terms in its quadrupole moment $q$.}
{We have identified a suitable class of frequency relations well fitting the observed data. These models imply for central compact object in \mbox{4U~1636-53} the mass $M\,=\,1.6\,$--$\,2.5M_{\sun}$, dimensionless angular momentum $j\,=\,0\,$--$\,0.4$, and quadrupole momentum $q\,=\,0\,$--$\,0.25$, with most preferred values $M\,\doteq\,1.77M_{\sun}$, $j\,\doteq\,0.05$, and $q\,\doteq\,0.003$.}
{The relationship implied for a particular case of so called total precession resonance between the Keplerian and the \dcp ~introduced in this paper resembles the twin peak QPOs observed in 4U 1636-53 with a $\chi^2\!\sim\!3\,{d.o.f.}$ which is about one order lower than $\chi^2$ reached by other theoretical relationships which we examine. Moreover, the position of 3/2 and 5/4 resonant points implied by the total precession relationship well coincides with frequencies given by the change of the rms-amplitude difference sign. Notice that if a resonance (in our opinion most likely present) is not considered, the total precession relation has a similar kinematic meaning as the periastron precession relation involved in the model of Stella and Vietri, but gives substantially better fit and lower neutron star mass.}

\authorrunning {Z. Stuchl\'{\i}k, G. T\"or\"ok, P. Bakala
}

\titlerunning{On a multi-resonant origin of high frequency QPOs in \mbox{4U~1636-53}}

\maketitle
   
\begin{figure*}
\begin{minipage}{1\hsize}
\begin{center}
\includegraphics[width=.8\textwidth]{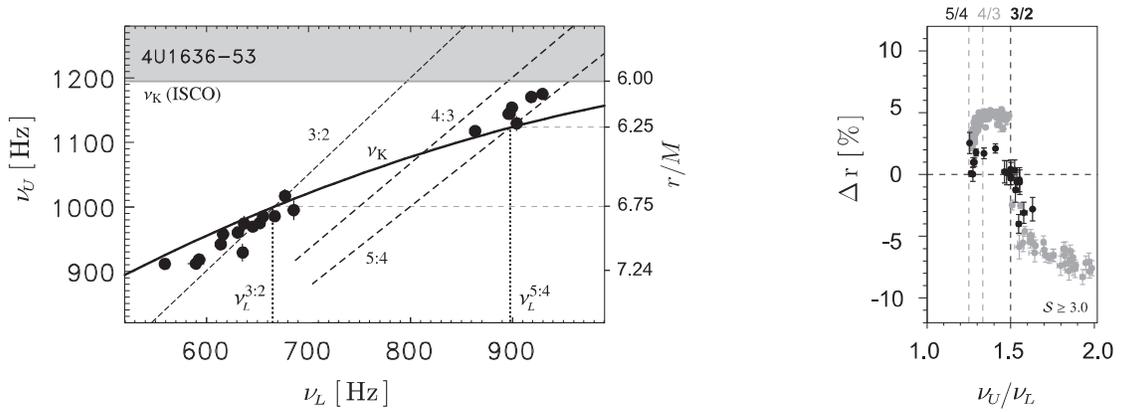}
\end{center}
\end{minipage}
\caption[t]
{Left \citep[from][]{tor-etal:2007}: The frequency correlation in the atoll source 4U 1636-53. Curve $\nu_\mathrm{K}$ determines the upper QPOs frequency following from the relativistic precession model \citep{ste-vie:1999:PHYRL:} under the consideration of the gravitational field described by the Schwarzschild metric with the central mass $M\!=\!1.84M_{\odot}$. 
The secondary vertical axes indicates the appropriate dimensionless radius. One should note that the assumed 5:4 resonance occurs very close ($0.25M$) to the \emph{innermost stable circular geodesic orbit}, i.e., near the expected inner edge of the accretion disc.~
Right \citep[from][]{tor-bar:2007}: The rms amplitude difference as a function of the frequency ratio in 4U 1636-53 (black points). Gray points correspond to the single detections of the lower (upper) QPOs and the absolute value of their rms amplitude. 
}
\label{figure:observation}
\end{figure*}
%
\section{Introduction}

Rossi X-ray Timing Explorer \citep[RXTE,][]{bra-etal:1993:AA} provides observations of the high frequency kilohertz QPOs in the X-ray fluxes from neutron-star binary systems  \citep[see,] [for a review]{kli:2006}.

Several models have been outlined to explain the kHz QPO frequencies, and it is mostly preferred that their origin is related to the orbital motion near the inner edge of an accretion disc. In particular, two ideas based on the strong gravity properties have been proposed. While \cite{ste-vie:1998:ASTRJ2:,ste-vie:1999:PHYRL:} introduced the ``Relativistic Precession Model'' in which the kHz QPOs represent direct manifestation of the modes of a relativistic epicyclic motion of blobs in the inner parts of the accretion disc, \cite{klu-abr:2000} proposed models based on resonant interaction between orbital and/or epicyclic modes related to non-linear oscillations of the accretion disc.

\cite{abr-etal:2003} noticed that ratio between the frequencies $\nuL$ and $\nuU$ of the lower and upper observed kHz QPO mode cluster in neutron-star sources usually close to ratios of small natural numbers, most often close to the value \mbox{$\nuU/\nuL\,=\,3/2$}.\footnote{
The same 3/2 ratio of QPO frequencies was first noticed in the GRO J1655-40 data by \cite{abr-klu:2001:AA} and later found in all other microquasars in which the twin QPOs have been detected \citep[see,][]{mcC-Rem:2004}.}
The ratio clustering was later confirmed by \cite{bel-etal:2005}. The question whether such a clustering represents clear argument supporting a resonance hypothesis remains to be subject of discussions. Nevertheless it was also suggested that, due to multipeaked distribution in the frequency ratio, more than one resonance could be at work if a resonant mechanism is involved in generating the neutron-star \citep{bel-etal:2005,tor-etal:2007} and black-hole \mbox{\citep{stu-tor:2005}} kHz QPOs.

\cite{tor-bar:2007} realized an interesting root-mean-squared-amplitude (rms amplitude) evolution in group of six neutron star atoll sources (namely 4U 1636-53, 4U 1608-52, 4U 1820-30, 4U 1735-44, 4U 1728-34 and 4U 0614+09) - the upper and lower QPO amplitudes equal each other when the source passes through a rational frequency ratio. Such a  behaviour
highly suggests the existence of an energy overflow between the upper and lower frequency mode typical for non-linear resonances. The effect seems to be related to the results of the frequency ratio distribution studies because the change of the rms amplitude difference sign occurs close to the same frequency ratio as those corresponding to the datapoints clustering.

\emph{The results of studies mentioned above indicate that for a given source the upper and lower QPO frequency can be traced through the whole observed range of frequencies but the probability to detect both QPOs simultaneously increases when the frequency ratio is close to the ratio of small natural numbers} \citep[namely 3/2, 4/3 and 5/4 in the case of six atoll sources,][]{tor-bar:2007}.

Figure \ref{figure:observation} illustrates the frequency correlation for the atoll source 4U 1636-53 together with its rms amplitude difference evolution.\footnote{The data \citep{bar-etal:2005,abr-etal:2005:rag} correspond to individual continuous segments of observation processed by standard shift--add technique \citep[see,][]{men-etal:1998}. Notice that some other methods, namely shift-add through all segments of data \citep[see,][]{bar-etal:2005}, provide more efficient analysis of the frequency correlation but do not keep the information about the probability of the detection (i.e., the frequency distribution).} For the $3/2$ and $5/4$ frequency ratio the two distinct clusters of datapoints are easy to recognize on the figure as well as the corresponding change of the rms amplitude difference sign.

In this paper we focus on 4U 1636-53 and explore the idea \citep{tor-etal:2007} that the two clusters of datapoints mentioned above result from two different instances of the same orbital resonance varying in resonant coefficients. 

\section{Orbital frequencies of geodesic motion close to rotating neutron stars}
\label{section:geodesic}

In Newtonian physics, a test particle orbits the central mass $M$ with the Keplerian angular velocity 
$\Omega^* = (GM/r^3)^{1/2}$
~and its, in general eliptic, trajectory is closed, i.e., related \emph{azimuthal (or {Keplerian})} $\nuK^*$, \emph{radial} $\nu^*_\mathrm{r}$, and \emph{vertical} $\nu^*_\theta$ orbital frequencies equal each other.

In the general relativistic description, necessary in the case of compact objects inluding neutron stars \citep[][]{mis-etal:1973, sha-teu:1986, gle:1997}, the trajectory of a test particle is not closed 
 and the frequencies of the azimuthal, radial and vertical ``quasielliptic'' orbital motion differ. For a given axially symetric spacetime with the line element
\begin{eqnarray}
\dd s^2 = \g_{tt}\,\dd t^2 + \g_{rr}\,\dd r^2 + \g_{\theta \theta}\,\dd \theta^2
             + \g_{\phi\phi}\,\dd \phi^2 + \g_{\phi t}\,\dd \phi\,\dd t+\nonumber\\
	                                 + \g_{t \phi}\,\dd t\,\dd\phi \,,~~~~~~~~~~~~~~~~~~~~~~~~~~~~~~~~~~~~~~~~~~~~~~~~~~~~~~~~~~~~~~~
\end{eqnarray}
the relevant  angular velocities read \citep[e.g.,][]{ali-gal:1981,oka-etal:1987,ste-vie:1998:ASTRJ2:,abr-joa:2003}\footnote{In the text above we mark the frequencies expressed in standard (SI) physical units by asterisk. Hencefort we already use the geometrical units ($c\,=\,1\,=\,G;~M\,=\,GM^*/c^2,~r\,=\,r^*,~t\,=\,ct^*$).} 
\begin{eqnarray}
\label{equation:omega}
\Omega_{\mathrm{K}}  &=& u^\phi/u^t,\\
\label{equation:epicyclic}
\omega^2_{\mathrm{i}} &=&
 \frac{(\g_{tt} +\Omega_\pm\,\g_{t\phi})^2}{2\g_{\mathrm{ii}}}\,
 \left(\pD{^2U}{\mathrm{i}^2}\right)_{\ell}\,,
\end{eqnarray}
where $\mathrm{i} \in (\mathrm{r},\theta)$ and $U$ is an \emph{efffective potential}
\begin{equation}
U(r,\theta,\ell) := \g^{tt} - 2\ell\,\g^{t\phi} + \ell^2\,\g^{\phi\phi},
\end{equation}
with $\ell$ denoting the specific angular momentum of the orbiting test particle
\begin{equation}
\ell = - u_\phi/u_t\,.
\end{equation}

In the following we consider Keplerian motion and $l\,=\,l_K(r,\theta)$. The motion is then described by the Keplerian frequency $\nu_K$ and radial and vertical epicyclic frequencies $\nur,~\nuv$.

Due to the inequality between the azimuthal and radial frequency, the eccentric orbits waltz at the \emph{periastron precession frequency} \citep[e.g.,][]{mis-etal:1973}
\begin{equation}
\label{equation:periastron}
\nuP=\nuK-\nu_\mathrm{r},
\end{equation}
and in addition the orbits tilted relative to the equatorial plane of the spinning central mass wobble at the \emph{nodal (often called Lense--Thirring) precession frequency} \citep[e.g.,][]{mis-etal:1973}
\begin{equation}
\label{equation:lensethirring}
\nuLT=\nuK-\nuv.
\end{equation}

The periastron precession frequency (\ref{equation:periastron}) corresponds to the period in which the test particle ``quasielliptic'' trajectory periastron oscillates and the nodal (Lense-Thirring) precession frequency (\ref{equation:lensethirring}) corresponds to the period in which the declination of the quasiellipse plane oscillates. Both the declination of the quasiellipse plane and position of the periastron then reach the initial state simultaneously in the period characterized by the frequency
\begin{equation}
\label{equation:BTS:def}
\nuBTS=\nuP-\nuLT=\nuv-\nur.
\end{equation}
Therefore, for the purposes of our paper, we call this frequency \emph{\dcp}. \footnote{Note that in general the repeating of initial periastron position and orbit declination do not guarantee repeating of the initial test particle position at the same time.}

\subsection{The Hartle-Thorne metric}
\label{section:methodI}

In this paper we describe the external neutron star spacetime using the Hartle-Thorne metric \citep{har-tho:1968} which represents the exact solution of vacuum Einstein field equations for the exterior of rigidly and relatively slowly rotating, stationary and axially symmetric body. The metric is given with accuracy
up to the second order terms in the body's dimensionless angular momentum$j = J/M^2$, and first order
in its dimensionless quadrupole moment $q =- Q/M^3$.
The explicit form of formulae (\ref{equation:omega}) and (\ref{equation:epicyclic}) derived by \cite{abr-joa:2003}, which we use in a slightly modified form, is given in Appendix \ref{appendix:hartle-thorn}.

\section{Testing the multiresonant hypothesis}
\subsection{Frequency identification}

Usually the $n:m$ orbital resonant models considering a non-linear resonance between Keplerian and/or epicyclic frequencies \citep[see, e.g.][]{abr-etal:2004} identify the resonant eigenfrequencies $\nuL^0,~\nuU^0$ as 
\begin{equation}
\label{equation:usualresonances}
\nul^0=\nur(r_{n:m}),~\nuu^0=\nu_\mathrm{v}(r_{n:m}),~~\nu_\mathrm{v}\in[\nuv,~\nuK]
\end{equation}
where $n,m$ are small natural numbers and $r_{n:m}$ is the radius fixed by the condition
\begin{equation}
\label{equation:standard}
\frac{\nu_\mathrm{v}(r_{n:m})}{\nur(r_{n:m})}=\frac{n}{m}.
\end{equation}

In the case of a considerably weak forced or parametric non-linear resonance \citep{lan-lif:1976}, the upper and lower observed QPO frequencies $\nuL$ and $\nuU$ are related to the resonant eigenfrequencies either directly
\begin{equation}
\nuL\doteq\nul^0,\quad \nuU\doteq\nuu^0,
\end{equation}
or as their linear combinations
\begin{equation}
\label{equation:combinations}
\nuL\doteq\alpha\nul^0,\quad \nuU\doteq\beta\nuu^0,
\end{equation}
where $\alpha$ and $\beta$ are small integral numbers. This property was utilized to estimate the spin of microquasars displaying constant twin peak QPOs from resonant models \citep{abr-klu:2001:AA,tor-etal:2005:AA}.

In general case of a system in a non-linear resonance, the observed frequencies differ from resonance eigenfrequencies by a frequency corrections proportional to the square of small dimensionless amplitudes \citep{lan-lif:1976}. It was shown \citep{abr-etal:2003:PASJ,reb:2004} that a resonance characterized by one pair of eigenfrequencies may reproduce the whole range of frequencies observed in a neutron star source. Later \cite{abr-etal:2005:rag} considered the idea of one eigenfrequency pair (so called resonant point in the frequency-frequency plane) common for a set of neutron star sources. They shown that for weakly coupled non-linear resonance the upper and lower frequency observed in a source should be linearly correlated. They also found that coefficients of linear fits well approximating individual sources are anticorrelated which was in a good accord to the theory they presented and justified the hypothesis of one eigenfrequency-pair. On the other hand this approach, incorporating certain difficulties (e.g., the extremely large extension of the observed frequency range), is not proved yet, and some observational facts like the multipeaked ratio distribution suggest that {more then one resonant points may be responsible for the almost linear observed frequency correlation}.

In next \emph{we focus on the hypothesis of more resonant points corresponding to different instances of one resonance and suppose that the observed frequencies are close to the resonance eigenfrequencies, i.e. that the observed frequency correlation follows the generic relation between resonant eigenfrequencies},
\begin{equation}
\label{equation:assumption}
\nuL\sim\nul^0,\quad \nuU\sim\nuu^0.
\end{equation}

For the whole applicable range of the internal angular momentum $j$ of the Hartle-Thorne spacetimes we checked that the ratio between the Keplerian (or vertical epicyclic) frequency and radial epicyclic frequency monotonically increases with decreasing radius $r$ whereas the Keplerian (vertical epicyclic) frequency increases \citep[see also][]{tor-stu:2005:AA}.

In other words, \emph{for the models (\ref{equation:usualresonances}) considering resonance between Keplerian (vertical epicyclic) frequency and radial epicyclic frequency satisfying relation (\ref{equation:assumption}), the ratio of observed frequencies should increase with increasing QPO frequency, but that is opposite to what is observed}.

However, the relations (\ref{equation:standard}--\ref{equation:combinations}) are not the only possible in the framework of resonance models. \citet{bur:2005:rag} discussed so called vertical precession resonance introduced in order to match the spin estimated from fits of the X-ray spectral continua for the microquasar \mbox{GRO~J1655-40}. The resonance should occur between the vertical epicyclic frequency and the periastron precession frequency fulfilling the relation
\begin{equation}
\label{equation:bursa}
\nul^0(r)=\nuP(r)=\nuK(r)-\nur(r),\quad\nuu^0(r)=\nuv(r),
\end{equation} 
for a particular choice of the resonant radius $r$ defined by the condition $\nu_\mathrm{u}\,=\,3/2\nu_\mathrm{l}$.

As noticed in \cite{tor-etal:2007}, for the Schwarzschild spacetime the relations (\ref{equation:bursa}) coincide with those following from the relativistic precession model:
\begin{equation}
\label{equation:stella}
\nul^0(r)=\nuP(r)=\nuK(r)-\nur(r),\quad\nuu^0(r)=\nuK(r).
\end{equation}
Opposite to the relations (\ref{equation:usualresonances}) the two relationships (\ref{equation:bursa},\ref{equation:stella}) as well as the other two relationships
\begin{eqnarray}
\label{equation:BTS2}
\nul^0(r)&=&\nuv(r)-\nur(r),\quad\quad\quad\nuu^0(r)=\nuv(r),\\
\label{equation:BTS}
\nul^0(r)&=&\nuBTS(r)=\nuv(r)-\nur,\quad\,\nuu^0(r)=\nuK(r)
\end{eqnarray}
imply the increase of $\nuu^0$ for increasing $\nul^0$.

In the following subsection we fit the QPO frequencies observed in \mbox{4U 1636-53} by the four different frequency relationship (\ref{equation:bursa}--\ref{equation:BTS}), testing the hypothesis that an appropriate resonance may be responsible for all the observed datapoints.

\begin{figure*}
\begin{minipage}{1\hsize}
\begin{center}
\includegraphics[width=1\textwidth]{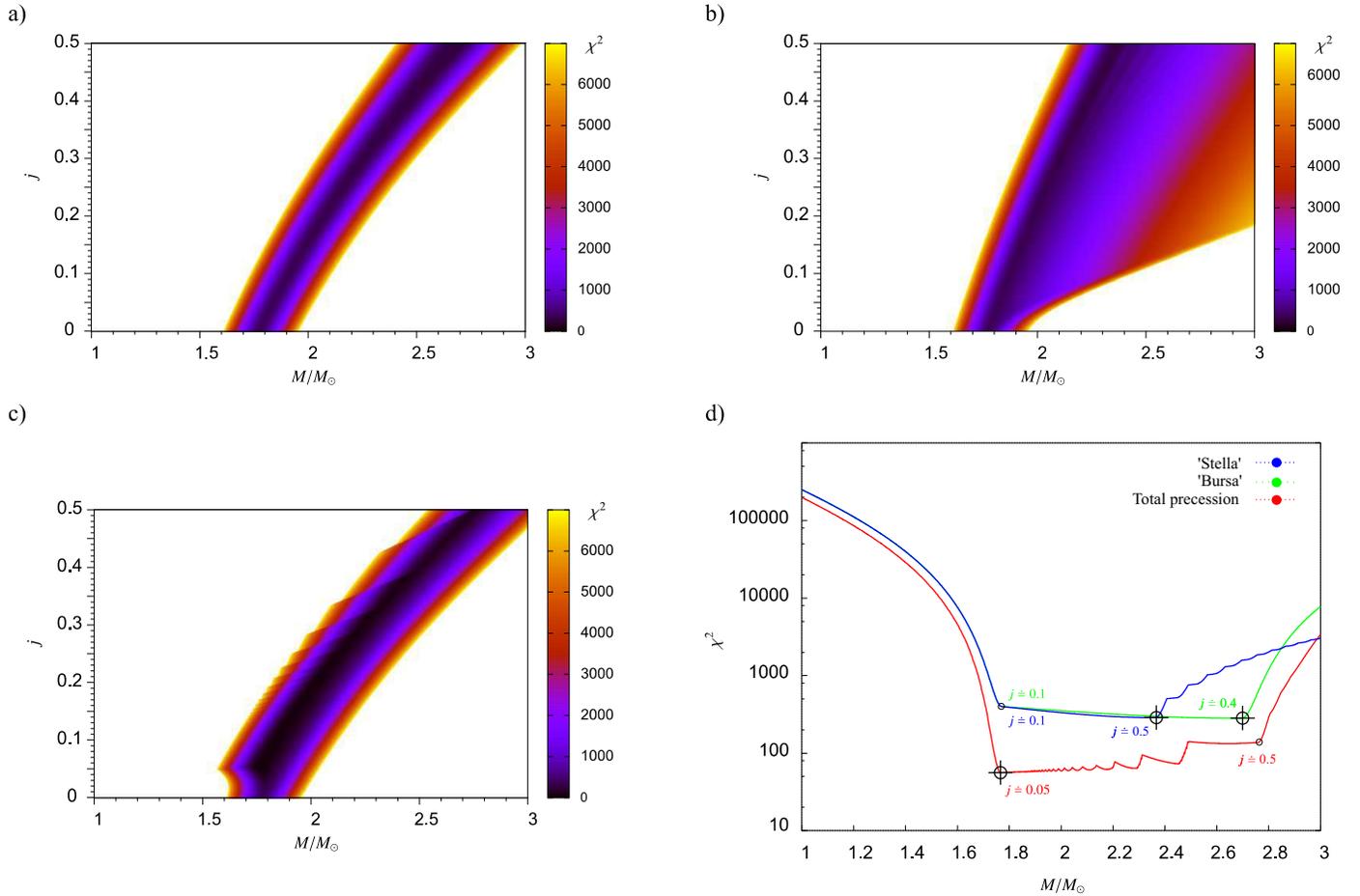}
\caption
{a) The inverse quality measure $\chi^2$ of the datapoints fits by relation (\ref{equation:stella})$\,$--$\,$``Stella'' as a function of the neutron star mass $M$ and the angular momentum $j$. The displayed individual values of $\chi^2$ correspond to the quadrupole momentum $q$ giving the best fit for the fixed pairs ($M,~j$). b)~The same for the relation (\ref{equation:bursa})$\,$--$\,$``Bursa''. c)~The same  but for the relation (\ref{equation:BTS})$\,$--$\,$ total precession. For all the three relations, the $\chi^2$ minima form ``valleys'' with rather sharp roots.
 d) Profiles of the individual valleys. The cross-hair marks denote the value of the global minima. For relations ``Stella'' and ``Bursa'' the value of $\chi^2$ local minimum at the bottom of the valley decreases with the increasing angular momentum $j$. On the other hand for the total precession relation it increases with the increasing angular momentum $j$.}
\label{figure:maps}
\end{center}
\end{minipage}
\end{figure*}

\subsection{Individual fits}
\label{section:fits}

The fits are realized in two steps. In order to obtain a rough scan we calculated frequency relations (\ref{equation:bursa}--\ref{equation:BTS}) in the Hartle--Thorne metric for the range of the mass $M\in\!1$--$3M_{\sun}$, the internal angular momentum  $j\in\!0$--$0.5$ and a physically meaningful quadrupole momentum $q$ with a step equivalent to the thousand points in all three quantities, i.e., four 3-dimensional maps each having $10^9$ points. Then, for each pair $(M,j)$, we keep the value of the quadrupole momentum $q$ which gives the lowest $\chi^2$ with respect to the observed datapoints. For the Schwarzschild spacetime ($q=j=0$), when relations (\ref{equation:bursa}--\ref{equation:BTS}) merge, the best fit is reached for the mass $M\,\doteq\,1.77M_{\sun}$, with a $\chi^2\doteq400\sim20\,d.o.f$. Because for rotating configurations the relation (\ref{equation:BTS2}) gives rather less interesting results, we do not consider this relation in the following.

Figures \ref{figure:maps} a, b, and c show the lowest $\chi^2$ associated with relations (\ref{equation:stella})$\,$--$\,$``Stella'', (\ref{equation:bursa})$\,$--$\,$``Bursa'', and (\ref{equation:BTS})$\,$--$\,$ total precession, for each pair ($M,j$).
Having a rough clue given by Figures \ref{figure:maps}a,b, and c we searched for local $\chi^2$ minima using the Marquardt--Levenberg non-linear least squares method \citep{mar:1963}. Particular minima we have found are denoted in Figure \ref{figure:maps}d.

The detailed analysis shown that both the relations (\ref{equation:bursa}) and (\ref{equation:stella}) match the observational data most likely for relatively high angular momentum close to $j\!\sim\!0.5$ and the central mass $M\!\sim\!2.4$--$2.8M_{\sun}$ (``Stella''), and $M\!\sim\!2.4M_{\sun}$ (``Bursa'') respectively, reaching the value of $\chi^2\!\sim\!15\,d.o.f.$, whereas the quadrupole momentum is rather close to the Kerr value $q\!=\!a^2\!=\!j^2$ $(q\!\sim\!0.23,\,0.25)$.

Relation (\ref{equation:BTS}, total precession) gives the best fit with the remarkable $\chi^2\!\sim\!3\,d.o.f.$ for angular momentum $j\sim\!0.05$ and central  $M\!\sim\!1.76\,M_{\sun}$, again with the quadrupole momentum close to the Kerr value $(q\!\sim\!0.029$)\footnote{Our results are thus compatible with expectation that gravitational field of settled down neutron stars can be well described by quasi-Kerr spacetime.}. For this relationship the quality of the fit is then very close to $\chi^2\!\sim\!3\,d.o.f.$ in rather large range of the central mass $M\!\sim1.76$--$1.84M_{\sun}$ and the angular momentum $j\!\sim0.05$--$0.1$.

\vspace{-1ex}

\section{Discussion and conclusions}

\begin{figure*}[t!]
\begin{minipage}{1\hsize}
\begin{center}
\includegraphics[width=1\textwidth]{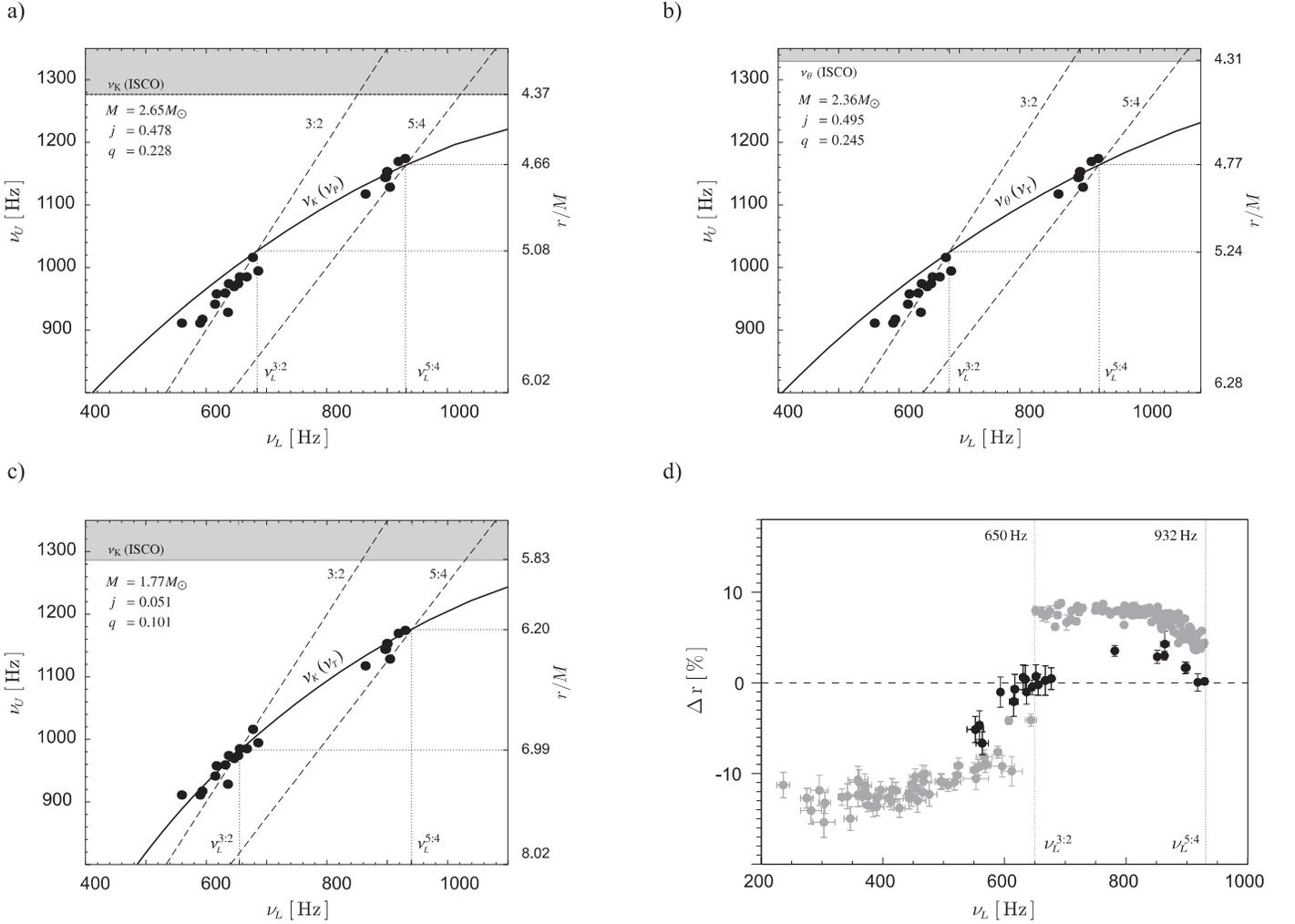}
\caption[t]
{
a) The best fit by relation (\ref{equation:stella})$\,$--$\,$``Stella''. 
b) The best fit by relation (\ref{equation:bursa})$\,$--$\,$``Bursa''. 
c) The best fit by relation (\ref{equation:BTS})$\,$--$\,$total precession. 
d) Adopted from \cite{tor-bar:2007}. The difference of the lower and upper QPOs rms amplitudes plotted as a function of the lower QPOs frequency (black points). Assuming zero amplitude of the missing QPOs mode, the quantity $\Delta\mathrm{r}$ is for single peaks (grey points) given by an absolute value of their rms amplitude. To plot $\Delta\mathrm{r}$ corresponding to the upper single QPO we used the linear frequency--frequency fit. Dotted vertical lines denote position of the qualitative change in the rms amplitude interrelationship. }
\label{figure:thelast}
\end{center}
\end{minipage}
\end{figure*}

Figures \ref{figure:thelast}a,b, and c show the best fits reached by relations (\ref{equation:bursa}--\ref{equation:BTS}). The relevant best fits properties are summarized in the Table \ref{table:fits}. Apparently, the identification of the lower kHz QPOs with \dcp~ and the upper kHz QPOs with the Keplerian frequency\footnote{As previously stressed, our paper focuses on the hypothesis of one resonance which occurs at different resonant points. The more general multiresonant idea, including the possibility of several resonances sharing the common radius, relevant for compact objects with special values of the angular momentum, is discussed in \cite{stu-etal:2007}.}
\begin{eqnarray}
\nul^0(r)=\nuBTS(r)=\nuv(r)-\nur(r),\quad\,\nuu^0(r)=\nuK(r)
\nonumber
\end{eqnarray}
~reaches a substantially better quality of the fit $(\chi^2\!\doteq\!3\,d.o.f.)$ then the other three possibilities. The corresponding value of the angular momentum $j\sim0.05$--$0.1$ and mass $M\!\sim\!1.8$ are not excluded by any study published so far \citep{str-mar:2001,gil-etal:2002} and this frequency identification appears more likely than the relationship corresponding to the relativistic precession model of Stella \& Vietri $(M\!\sim\!2.4$--$2.8M_{\sun},~j\!\sim\!0.4$--$0.5)$.

The total precession frequency $\nuBTS$ given by (\ref{equation:BTS:def}) corresponds to the period in which the declination of the free test particle quasiellipse plane and the periastron reach simultaneously the initial state. Consideration of a hot spot with characteristic frequencies (\ref{equation:BTS}) may represent a kinematic QPOs model very close (but obviously not identical) to those of Stella \& Vietri.\footnote{One may also argue that if the light curve is somehow simultaneously modulated by the periastron and Lense-Thirring precession, the beat frequency should appear in its power spectra.}

However, the observed ratio clustering and rms amplitude difference behaviour suggest the  existence of a resonance between the lower and upper QPOs modes. Notice also that the position of the 3/2 and 5/4 resonant points implied by the total precession relationship (i.e., intersections of the best datapoints fit with reference 3/2 and 5/4 lines) well coincide with frequencies given by the change of the rms-amplitude difference sign (see Figure \ref{figure:thelast} and Table \ref{table:fits}). Therefore, our results indicate that the resonance may occur between the Keplerian frequency of the trajectory and the \dcp~ corresponding to the periodicity of the trajectory shape.\footnote{For the perfect free particle motion, if the Keplerian and total precession frequency form rational fractions, the trajectory is selfrepeating (i.e., closed).} The concrete physical mechanism of such a resonance remains to be the subject of a future and rather larger research since, beyond the hotspot interpretation, the Keplerian and \dcp~ may also correspond to some disk oscillation modes.

The results we have obtained so far indicate that Keplerian and \dcp~ (\ref{equation:BTS}) match the kHz QPOs frequencies very well at least in several atoll sources \citep[][in preparation]{bak-etal:2007}.

On the other hand, the applicability of the relation (\ref{equation:BTS}) to the Galactic microquasar sources poses an open question because it implies the central black hole spin similar to those of Stella \& Vietri ($a\!\sim\!0.3$--$0.5$), which in the case of microquasar \mbox{GRS~1915+105} contradicts the recent results of a fitting the spectral continua \citep[$a\!\sim\!0.7$--$1$,][]{McCli-etal:2006:ASTRJ2:,Mid-etal:2006:MONNR:}.

\begin{table*}[t!]
\begin{center}
 \caption{The properties of best fits depicted in Figure \ref{figure:thelast}a,b, and c (in the case of the totall precession we display also properties of the fit for the angular momentum $j\,\doteq\,0.1$). The uncertainities in the fit parameters are the standard $\chi^2+1$ errors. Quantity $\Delta q$ characterizes the deviation of the quadrupole momentum from the value corresponding to the Kerr spacetime. Quantities $\Delta\nuL$ characterize the deviation of best fit intersections with reference 3/2 and 5/4 lines from the value given by the qualitative change in the rms amplitudes behaviour.}{\label{Table1}} 
\label{table:fits} 
\renewcommand{\arraystretch}{1.2}
\begin{tabular}{c c  c c c c c c c c c c c c c}
\hline
\multicolumn{2}{c}{Model}&\multicolumn{5}{c}{Best fit}&&\multicolumn{7}{c}{Resonant points}\\
$\nuL$&$\nuU$&$M$&$j$&$q$&$\Delta q $&{\large{$\frac{\chi^2}{d.o.f.}$}}&&\large$\frac{r_{3:2}}{M}$&\large$\frac{\nuL^{3/2}}{\mathrm{Hz}}$&\large$\frac{r_{5:4}}{M}$&\large$\frac{\nuL^{5/4}}{\mathrm{Hz}}$&\large$\frac{\Delta\nuL^{3/2}}{\mathrm{Hz}}$&\large$\frac{\Delta\nuL^{5/4}}{\mathrm{Hz}}$&\large$\frac{\,\overline{\Delta\nuL}\,}{\mathrm{Hz}}$\\\hline
\multicolumn{2}{c}{``Stella''}&\multicolumn{5}{c}{~}&&\multicolumn{7}{c}{~}\\
$\nuP$&$\nuK$&
2.65$\pm0.20$&0.478$$$\pm0.099$&0.228$\pm0.096$&0.000&15&&5.08&684&4.66&931&34&1&17.5\\
\hline
\multicolumn{2}{c}{``Bursa''}&\multicolumn{5}{c}{~}&&\multicolumn{7}{c}{~}\\
$\nuP$&$\nuv$&
2.36$\pm0.01$&0.495$\pm0.005$&0.245$\pm0.004$&0.000&15&&5.24&683&4.77&931&33&1&17.0\\
\hline
\multicolumn{2}{c}{Total}&\multicolumn{5}{c}{~}&&\multicolumn{7}{c}{~}\\
\multicolumn{2}{c}{precession}&\multicolumn{5}{c}{~}&&\multicolumn{7}{c}{~}\\
$\nuBTS$&$\nuK$&
1.77$\pm0.07$&0.051$\pm0.044$&0.003$\pm0.009$&0.0003&3&&6.99&655&6.20&940&5&8&6.5\\
\hline
\multicolumn{2}{c}{Total}&\multicolumn{5}{c}{~}&&\multicolumn{7}{c}{~}\\
\multicolumn{2}{c}{precession}&\multicolumn{5}{c}{~}&&\multicolumn{7}{c}{~}\\
$\nuBTS$&$\nuK$&
1.84$\pm0.08$&0.101$\pm0.044$&0.011$\pm0.009$&0.0001&3&&6.8&653&6.01&941&3&9&6.0\\
\hline
\end{tabular}
\end{center}
\medskip
\end{table*}

\section*{Acknowledgments}
The authors have been supported by the Czech grants MSM~4781305903 (ZS and GT) and LC~06014 (PB).


\onecolumn
\begin{appendix}

\section{Formulae for orbital geodesic frequencies in the Hartle--Thorne metric}
\label{appendix:hartle-thorn}

After \citet{abr-joa:2003}, the Keplerian orbital angular velocity and the radial and vertical epicyclic angular velocities can be expressed in terms of the Hartle-Thorne metric parameters $M,~j,~,q$ in the following form.
\medskip

The angular velocity for corotating
 circular particle orbits reads

\begin{equation}
\Omega_\K  = u^\phi/u^t = 
\frac{M^{1/2}}{r^{3/2}}
               \,\left[ 1 
               -j\,\frac{M^{3/2}}{r^{3/2}} 
               + j^2\,F^{~\Omega}_1(r) + q\,F_2^{~\Omega}(r) \right],
\end{equation}
where
\begin{eqnarray}
F_1^{~\Omega}(r)    &=& \left[ 48\,M^7 - 80\,M^6\,r + 4\,M^5\,r^2 - 
                 18\,M^4\,r^3 + 40\,M^3\,r^4 + 10\,M^2\,r^5 \right.\\
	     & & {} \left. + 15\,M\,r^6 -  15\,r^7 \right] \,  
                 (16\,M^2\,\left( r - 2M \right) \,r^4)^{-1} + H(r) \non\\
   F_2^{~\Omega}(r)    &=& \frac{5\,\left( 6\,M^4 - 8\,M^3\,r - 2\,M^2\,r^2 - 
                 3\,M\,r^3 + 3\,r^4 \right) }{16\,M^2\,
                  \left( r - 2M \right) \,r} - H(r) \\[1ex]
   H(r)      &=& \frac{15\,\left( r^3 - 2\,M^3\right)}{32\,M^3} \,
                 \ln \left(\frac{r}{r - 2M}\right)\,.
\end{eqnarray}

\noindent
The epicyclic frequencies of circular geodesic motion are given by formulae
\begin{eqnarray}
{\omega}_r^2      &=& \frac{M (r - 6M)}  {r^{4}} \, 
          \left[ 1 
          +j\, H_1(r) - j^2\, H_2(r) - q\, H_3(r)\right] \\
   {\omega}_\theta^2 &=& \frac{M} {r^{3}} \, 
          \left[ 1 
          - j\, I_1(r) + j^2\, I_2(r) + q\, I_3(r) \right] 
\end{eqnarray}
where
\begin{eqnarray}
 H_1(r) &=& \frac{6\,M^{3/2}\,(r + 2M)}{r^{3/2}\,(r - 6M)} \\
   H_2(r) &=& \left[8 M^2\,r^4\,(r - 2M)\,(r - 6M) \right]^{-1} 
              \left[384 M^8 -720 M^7 r -112 M^6 r^2 -76 M^5 r^3\right.\non \\
          & & \left. {} - 138 M^4 r^4 - 130 M^3 r^5 + 635 M^2 r^6 
                     - 375 M r^7 + 60 r^8 \right] + J(r)  \\[1ex]
   H_3(r) &=& \frac{5\,( 48 M^5 + 30 M^4 r + 26 M^3 r^2 - 127 M^2 r^3 
               + 75 M r^4 - 12 r^5)}{8 M^2\,r\,(r - 2M)\,(r - 6M)} -J(r)\\
   I_1(r)~&=& \frac{6\,M^{3/2}}{r^{3/2}} \\
   I_2(r)~&=& \left[8 M^2\,r^4\,(r - 2M) \right]^{-1} 
              \left[48 M^7 - 224 M^6 r + 28 M^5 r^2 \right. \nonumber \\
          & & \left. {} + 6 M^4 r^3 - 170 M^3 r^4 + 295 M^2 r^5 
                     - 165 M r^6 + 30 r^7 \right] - K(r)  \\[1ex]
   I_3(r)~&=& \frac{5\,\left( 6 M^4 + 34 M^3 r - 59 M^2 r^2 + 33 M r^3 
                     - 6 r^4 \right)}{8 M^2\,r\,(r - 2M)} + K(r),  
\end{eqnarray}
with
\begin{eqnarray}                     
   J(r)~~\,&=& \frac{15 r\,(r - 2M)\,(2 M^2 + 13 M r - 4 r^2)}{16 M^3\,(r - 6M)}
              \ln\,\left(\frac{r}{r - 2M}\right) \\
   K(r)~~&=& \frac{15 \,(2r - M)\,(r - 2M)^2}{16 M^3}
              \ln\,\left(\frac{r}{r - 2M}\right).
\end{eqnarray}

\noindent
For completness, the relation determining the marginally stable circular geodesic reads
\begin{equation}
r_{\mathrm{ms}} = 6\,M\,\left[ 1 
- j \,\frac{2}{3} \sqrt{\frac{2}{3}} + 
                  j^2\,\left( \frac{251647}{2592} 
                 - 240\,\ln \,\frac{3}{2} \right)
                 + q\,\left( -\,\frac{9325}{96}  
                 + 240\,\ln \,\frac{3}{2} \right) \right].
\end{equation}
\end{appendix}


\end{document}